\documentclass[% twocolumn,
aps,nofootinbib,
showpacs,showkeys,preprint
tightenlines,preprintnumbers,] {revtex4}

\usepackage{epsf,epsfig,subfigure,%axodraw,
graphicx,amsmath,amssymb}
\usepackage{color}

\newcommand{\ie}{{\it i.e.~}}
\newcommand{\etal}{{\it et al.\,}}

\usepackage{color}
\newcommand{\dis}[1]{\begin{equation}\begin{split}#1\end{split}\end{equation}} 

\newcommand{\qslash}{q\hskip -0.15cm\slash}
\newcommand{\kslash}{k\hskip -0.15cm\slash}
\newcommand{\pslash}{p\hskip -0.15cm\slash}
\newcommand{\Pslash}{P\hskip -0.22cm\slash}

\newcommand{\aslash}{a\hskip -0.15cm\slash}
\newcommand{\bslash}{b\hskip -0.15cm\slash}
\newcommand{\cslash}{c\hskip -0.15cm\slash}
\newcommand{\dslash}{d\hskip -0.15cm\slash}

\newcommand{\bfpe}{|{\bf p}_e'|}

\newcommand{\bfpa}{|{\bf P}_A'|}
\newcommand{\yunit}{{\rm Events/t\cdot y\cdot keV}}

\newcommand{\gev}{\,\textrm{GeV}}
\newcommand{\meV}{\,\mathrm{MeV}}
\newcommand{\keV}{\,\mathrm{keV}}
\newcommand{\eV}{\,\mathrm{eV}}

\def\ie{{\it i.e.~}}
\def\etal{{\it et al.\,}}

\def\sw0{{$\sin^2\theta_W^0$}}

\def\Nf2{{\bf N_{[2]}}}

\def\E6{{\rm E_6}}

\def\EE8{{\rm E_8\times E_8'}}

\begin{document}

\draft

\title{\bf Electromagnetic properties of neutrinos from  scattering on bound electrons in atom}

\author{ Junu Jeong$^{(1)}$, Jihn E.  Kim$^{(2)}$, and Sungwoo Youn$^{(1)}$} 
\address{
$^{(1)}$Center for Axion and Precision Physics Research, IBS, Daejeon 34051, Republic of Korea\\
$^{(2)}$Department of Physics, Kyung Hee University, 26 Gyungheedaero, Dongdaemun-Gu, Seoul 02447, Republic of Korea }

\begin{abstract} 
We consider the effects of bound atomic electrons scattered by solar neutrinos due to the electromagnetic properties of neutrinos. This necessiate considering the recoil of atomic nucleus, which should be considered in the momentum conservation, but that effect to the energy conservation is negligible. This effect changes the kinematic behavior of the scattered electron compared to that scattered on free electrons. We apply this effect to the recent  XENON1T data, but the bounds obtained from this is not very restrictive.   We obtained the bounds: the (transition) magnetic moment   $|f_{\alpha\beta}|\le 0.86\times 10^{-7}$ (times the electron Bohr magneton) and the charge radius  $|\tilde{r}|< 4.30\times 10^{-17\,}{\rm cm}$. For a non-vanishing millicharge ($\varepsilon$), the allowed bound is shown in the $\tilde{r}^2-\varepsilon$ plane.
 
\end{abstract}
\maketitle

%%%%%%%%%%%%%%%%%%%%%%%%%%%%%
%%%%%%%%%%%%%%%%%%%%%%%%%%
\section{Introduction}\label{sec:Introduction}
  
Ever since the discovery of neutrinos, it has been a great challenge to find the electromagnetic(EM) properties of neutrinos \cite{Pauli30,Reines57,LeeTD63}, which continued in the gauge theory era \cite{Clark73,Kim74,Kim78,Okun86,Barbieri87}. To have the EM properties of neutrinos at the observable level, particles in the beyond-the-standard-model(BSM) had to be introduced \cite{Kim76,Yanagida88,Giunti15,KimMoscow19}. Two relevant EM properties of neutrinos are the magnetic moment(MM) and the charge radius(CR).  The  MM, coupling to the EM field strength, is gauge invariant. On the other hand, the CR, coupling to the EM field itself, is not gauge invariant. Gauge invariance is needed to maintain the renormalizability. Therefore, if the probing wave length in the effective theory is much larger than the cutoff length-scale in the renormalization, as in the Fermi weak interaction much above the electroweak length-scale $10^{-16\,}$cm, CR can be also considered as a useful physical parameter. The cutoff energy scale for the CR is the mass of a heavy BSM particle whose exchange produces the relevant charge radius. In this paper, we take this viewpoint considering the electromagnetic properties of neutrinos.

The published experimental limit on the MM of $\nu_e$ in units of the electron Bohr magneton is \cite{BdNuMagMom17}, $
|f|\lesssim 2.8\times 10^{-11}$,
and the limit on the squared CR, $\tilde{r}^2$ is \cite{BdNuChargeR10}, $ 
\tilde{r}^2=  [-2.1, +3.3] \times 10^{-32}{\rm cm^2} $.
The standard model(SM) prediction \cite{Fujikawa80} on the neutrino MM is much smaller than the upper limit presented above, by a factor of  O($10^{-8}$).  Note that the millicharge limit of neutron is O($10^{-21\,}$cm) \cite{ChargeNeutron88}, which however cannot be directly used for a limit on the millicharges of neutrinos.

Recently, the XENON group considered the possibility of MMs of solar neutrinos for the excess events in their data,  for an exposure of 0.65 tonne-year with an extremely low background rate of $76\pm 2(\rm stat.)\,$(Events)/${\rm (t\cdot y\cdot keV)}$ \cite{Xenon1T20}.  The plot on these excess events starts around the electron recoil energy near 2--3\,keV, and ends around 30\,keV as shown in Fig. \ref{fig:FittoEvents}\,(a).
Out of the fitted 42,179.4 events, the estimated SM background by the solar neutrinos is 220.8 events \cite{Xenon1T20}, only 0.52\%. For estimating non-vanishing electromagnetic properties of solar neutrinos, therefore, one may neglect the SM background and associate the bulk of  42,179.4 events with the electromagnetic properties of neutrinos in the scattering process. In this way, Ref. \cite{Xenon1T20} obtained the bound on the neutrino MM, $ [1.4,2.9]\times 10^{-11}$ times the electron Bohr magneton. But, the assumed cross section by the XENON group has the reciprocal dependence on the recoil electron energy. This reciprocal dependence is obtained by assuming the two-body scattering \cite{Vogel89}, $\nu_e+e$, which may not be a correct method. The reason is that in kicking out an electron from a heavy atom, the atomic nucleus can carry away some momentum and the momentum conservation used in the $\nu_e+e_{\rm in\,atom}\to \nu_e'+e'$ is not exact.  

\color{black}
  In Sec. \ref{sec:XenonData}, we express the  XENON1T unit ``$\yunit$'' in terms of the cross section unit ``$\meV^3$'' to compare with the theoretical prediction.   In Sec. \ref{sec:Cross}, we discuss the possibility of including electromagnetic properties of neutrinos in the scattering process.   Here, the three-body phase space with a non-relativistic atom is discussed.  In Subsec. \ref{subsec:Mili}, we briefly comment on the possibility of  neutrino milli-charges. In Sec. \ref{sec:Fitting}, from XENON1T data we obtain the CR and millicharge bounds of electron-type neutrino. Section \ref{sec:Conclusion} is a brief summary on the EM properties of neutrinos. In Appendix \ref{AppA}, we estimate the parameters used in the calculation.

%%%%%%%
%%%%%%
\section{Event Rate and Cross Section in the XENON1T Data}\label{sec:XenonData}

%%%%%%%%%%%%%%%%%%
\begin{figure}[!h]
 \includegraphics[height=0.27\textwidth]{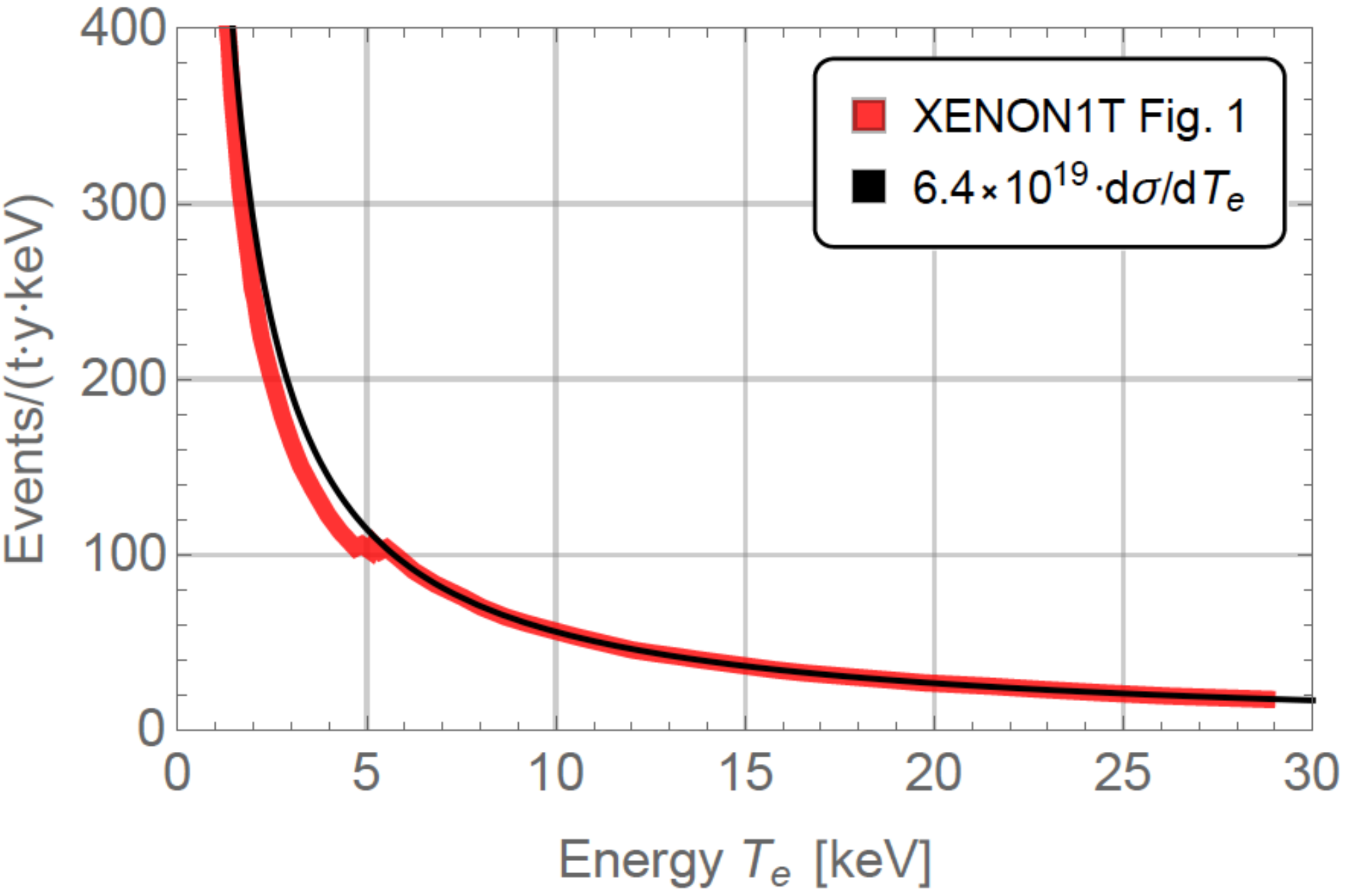}
\hskip 0.5cm
 \includegraphics[height=0.27\textwidth]{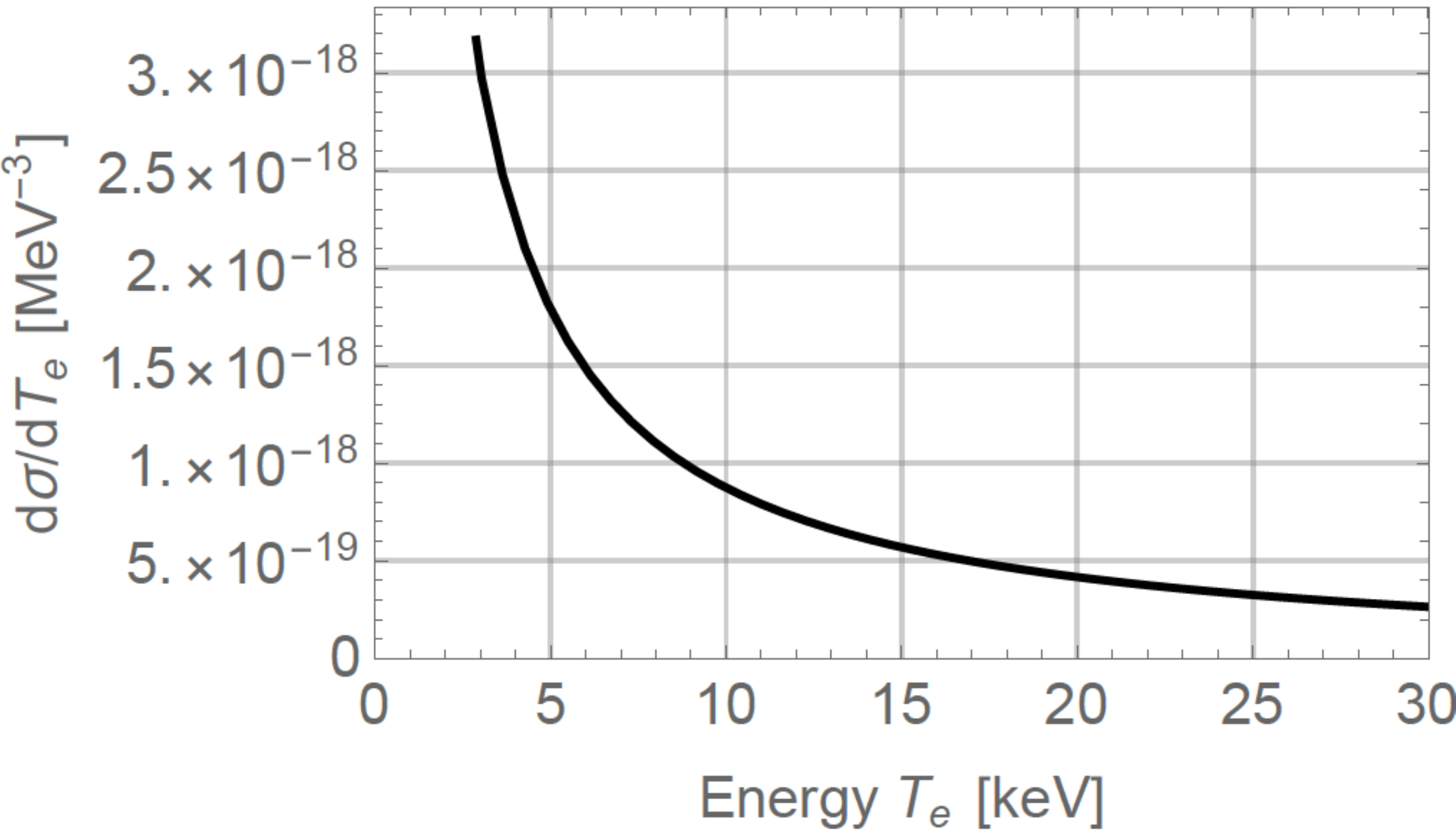} 
 \\~~ \\
 ~\hskip 1cm(a)\hskip 9cm (b)
\caption{(a) Cross section vs. the data curve for $\mu_\nu=7\times 10^{-11}\mu_{\rm B}$ reproducing the RHS panel of Fig. 1 of \cite{Xenon1T20}, and (b) a conversion of the event rate to the cross section.}
\label{fig:FittoEvents}
\end{figure}
%%%%%%%%%%%%%%%%

The XENON group presents data in units of $\yunit$.
 To convert it to cross section, we use the right-hand side (RHS) panel of Fig. 1 of \cite{Xenon1T20}, and followed  their analysis using the neutrino magnetic moment   $\mu_\nu=7\times 10^{-11}\mu_{\rm B}$ and the solar $pp$ neutrino flux of \cite{Bahcall04}.  They used  the inverse-$T_e$ rule presented in \cite{Vogel89}.  In our paper here, 
the $pp$ neutrino energy flux of \cite{Bahcall04} in the region, $E_\nu\ge 4\,\meV$, is fitted by $F^{\rm solar}_{\rm pp}(E)$.
The analytic form for the $pp$-flux function is presented in \cite{Vitagliano17},
\begin{equation}
    F^{\rm solar}_{pp}(E) = P_{0} E^{2} (Q + m_{e} - E) \sqrt{(Q + m_{e} - E)^{2} - m_{e}^{2}},
\end{equation}
where $Q \approx 423.41\,\textrm{keV}$ is the end point energy of neutrino, and $m_{e} \approx 0.511\,\textrm{MeV}$ is the electron mass.
$P_{0} = 188.143\,\textrm{MeV}^{-5}$ is the normalization constant.
From this form, some averages of energy functions are
\begin{equation}
\begin{split}
    \langle E^{-2} \rangle =&~ 30.0256\,\textrm{MeV}^{-2}, \\
    \langle E^{-1} \rangle =&~  4.46821\,\textrm{MeV}^{-1}, \\
    \langle E^{1} \rangle =&~ 0.267946\,\textrm{MeV}^{1}, \\
    \langle E^{2} \rangle =&~  0.0794926\,\textrm{MeV}^{2}, \\
    \langle E^{3} \rangle =&~  0.0251769\,\textrm{MeV}^{3}, \\
    \langle \ln{E/Q_{\rm min}} &\rangle = -0.528546. \\
\end{split}
\end{equation}

Figure \ref{fig:FittoEvents}\,(a) is the overlap of the cross section and the RHS panel data of Fig. 1 of Ref. \cite{Xenon1T20}. Here, we used $Q_{\rm min}$ as the minimum of $\bfpe^2/(2m_e)$.  Figure \ref{fig:FittoEvents}\,(b) is the cross section $d\sigma/dT_e$ in the MeV$^{-3}$ units, obtained from the XENON1T data using {\it  the  inverse-$T_e$ rule}.  Then, our calculation of $d\sigma/dT_e$ resulting from some electromagnetic properties of neutrinos, {\it not using the  inverse-$T_e$ rule but using our formula}, will be compared with Fig. \ref{fig:FittoEvents}\,(b).

%%%%%%%
  %%%%%%
\section{Cross Sections with Electromagnetic Form Factors of Neutrino}\label{sec:Cross}
%%%%%%%
 The photon vertex does not change the chirality. For the mass and MM, therefore, a chirality change must take place. So, the MM can arise by attaching a photon vertex in the mass generating diagrams.  For the CR, it is even simpler because the CR has the same chiral property as that given by the EM coupling. For the neutrino, the gauge coupling is expressed as the form factor $F^\nu_{1}(q^2)$ with $F^\nu_{1}(0)=0$. The CR is defined by the first term in the expansion of  $F^\nu_{1}(q^2)$ in terms of momentum transfer $q^2$. As noted in the beginning, we use the charge radius for the scale $q^2\ll$(cutoff scale)$^2$.   
In addition, if a (almost) massless extra photon (ex-photon) in the BSM sector  is present,  there is a possibility that neutrinos can carry millicharges \cite{Holdom86,ParkJC07}. The chiral property due to millicharges is the same as that due to the charge radius. The vector couplings of the charge radii and millicharges of neutrinos can mix with the $G_F$-order SM couplings, but the interference between  these vector and axial-vector interactions and the SM amplitude has not played an important role because of the tiny contribution in the data of XENON1T if it were from solar neutrinos.

%%%%%%
\subsection{SM cross section and beyond}\label{subsec:SM} 
%%%%%%%%%
The effective interaction in the SM for the electron-type neutrino scattering on an electron is given by the interaction \cite{tHooft71,KimRMP81},
\begin{eqnarray}
\frac{G_F}{\sqrt2}\bar{\nu}_e\gamma^\mu(1+\gamma_5)\nu_e\,
\bar{e}\gamma_\mu(g_V+g_A\gamma_5)e.\label{eq:SMcoupling}
\end{eqnarray}
When we consider the neutrino scattering, kicking out an electron from an atom with atomic number $Z$, the following six-fermion interaction with parameter  $M_{\rm eff}$  can be considered,  
\dis{
\frac{1}{M_{\rm eff}^3}\cdot\frac{G_F}{\sqrt2}\bar{\nu}_e\gamma^\mu(1+\gamma_5)\nu_e\,
\bar{e}\gamma_\mu(g_V+g_A\gamma_5)e\,\overline{A} A,\label{eq:Effatom}
}
where $A$ is treated as a  quantum field with $Z-1$ atomic electrons.\footnote{$A$ may be considered as a boson, which gives the same result.}
A strategy to estimate $M_{\rm eff}$ for four-fermion interaction is given in Appendix \ref{AppA}. We interpret  $M_{\rm eff}$ as representing the electromagnetic process occurring in the atomic orbits.  For four-fermion interaction, it is a kind of definition for the atomic process. For six-fermion interaction of (\ref{eq:Effatom}), we use the same  $M_{\rm eff}$   determined in Appendix \ref{AppA}.

In the SM, we have
\dis{
g_V=\frac12+2\sin^2\theta_W,~g_A=\frac12,
}
where $\sin^2\theta_W\simeq 0.231$ \cite{sin2LHC18,KimRMP81} is the weak mixing angle, and  $e^2=g^2\sin^2\theta_W$.
The EM couplings of neutrinos are 
\dis{
 \overline{\nu}_\beta\gamma^\mu e F_{1\,\alpha\beta}^\nu \nu_\alpha A^{\rm em}_\mu +  \overline{\nu}_\beta\sigma^{\mu\nu} \frac{ef_{\alpha\beta}}{2M}\nu_\alpha (q_\mu A^{\rm em}_{\nu}-q_\nu A^{\rm em}_{\mu})\label{eq:EMcoupling}
}
where $F_{1\,\alpha\beta}$ and $\frac{f_{\alpha\beta}}{2M}$ are matrices between weak eigenstates of neutrinos, $\nu_\beta(k_\nu')$ and $\nu_\alpha(k_\nu)$. $f_{\alpha\beta}$ is the MM of neutrino in units of the Bohr magneton of the mass $M$ particle. For $\alpha=\beta$, Eqs. (\ref{eq:SMcoupling}) and (\ref{eq:EMcoupling}) combine, and in particular for the electron-type neutrino $\alpha=\beta$ we have
\dis{
 \frac{G_F}{\sqrt2}\bar{\nu}_e\gamma^\mu(1+\gamma_5)\nu_e\,
\bar{e}\left(\left[g_V+ \frac{F_1^{\nu_e}}{\sqrt{2}G_F q^2}\right]\gamma_\mu+g_A\gamma_\mu\gamma_5\right)e.\label{eq:Combine}
}
The SM neutrino $\nu_\alpha$ is a two-component spinor. Therefore, in the cross section calculation the spin sums for a Majorana neutrino is $\frac12$ of that of a Dirac neutrino, which will be taken into account.

For calculating scattering cross section  on electrons bound in the orbits of a rest nucleus,  the wave function of electrons in the electron cloud around the nucleus must be considered. Let  the uncertainty of the outgoing electron position be  $\Delta r$. Then, the uncertainty of the momentum is  $\Delta p\approx 1/\Delta r$. We consider the recoil energy  greater than 2.5 keV (glimpsing Fig.  \ref{fig:FittoEvents}\,(a) ), corresponding to   to $\Delta r <3.9\times  10^{-9}$cm with  $\Delta p=\sqrt{2m_e T_e}$ for the electron recoil energy $T_e$. But, in the neutrino scattering with the incoming solar neutrino energy less than   $ 0.45\,\meV$ \cite{Bahcall04}, the uncertainty of neutrino position is less than $4.38\times 10^{-11}$cm which is much shorter than the orbit radius of the $n$-th level of Xenon, $10^{-10}n$\,cm. So, in the neutrino scattering we can safely use the two-body scattering formula as described in Appendix \ref{AppA}.

The $3\to n$  transition rate is
\dis{
Rate=(2\pi)^{4-3n}\frac{d^3{\bf p}_1'\cdots d^3{\bf p}_n' }{2E_12E_2 2E_32E_1'2E_2'\cdots 2E_n'} \frac{\delta^4(p_1+p_2+p_3-p_1'\cdots-p_n')}{V^2}
|\langle p_1'\cdots p_n'|T|p_1p_2p_3\rangle |^2\label{eq:3nRate}
}
from which we calculate the three body scattering cross section. A flux of incoming particles 1 (neutrino in our case) scatter on two particles 2 (electron in our case) and 3 (atom in our case). Thus, the cross section consists of two parts, one with a flux factor $1/|{\bf v}_1-{\bf v}_2|$ and the other $1/|{\bf v}_1-{\bf v}_3|$.
Taking ${\bf v}_3=0$, the neutrino-electron scattering is
\dis{
d\sigma=\frac{(2\pi)^{4-9}}{|{\bf v}_1-{\bf v}_2|}\frac{d^3{\bf p}_1'd^3{\bf p}_2'  d^3{\bf p}_3' }{2E_12E_2 2E_32E_1'2E_2'2E_3'} \frac{\delta^4(\sum p_i-\sum p_f')}{V}
|\langle p_1'p_2' p_3'|T|p_1p_2p_3\rangle |^2\label{eq:3scattT}
}
where we can take $V$ as the atomic volume to which the flux of neutrinos sweep. For $1/V$, the sum $B$ in Eq. (\ref{eq:Sumsn}) is used. The $T$-matrix squared in Eq. (\ref{eq:3scattT}) is summed over the spins
\dis{
\frac{1}{(2s_e+1)(2s_\nu+1)}\sum_{s_\nu,s_e}\sum_{s'_\nu,s'_e}|{\rm Amp}|^2=  (|T|^2_{\rm Q}).
}
If we consider the MM interaction also, then we will have the factor  $(|T|^2_{\rm MM} +|T|^2_{\rm Q})$ instead of $(|T|^2_{\rm Q})$. For the volume $V$, we use the Xenon volume for each principal quantum number $n$ separately. The $n$-th shell volume is $V=(4\pi /3)(na_Z)^3$ where $a_Z$ is the K-shell radius of Xenon atom.  Then, for the $\nu_\alpha(k_\nu)+e(p_e) +A(P_A)\to \nu'_\beta(k_\nu')+e'(p_e') +A'(P_A')
$ scattering, Eq. (\ref{eq:3scattT}) becomes
\dis{
d\sigma=\frac{3Z^3\alpha^3_{\rm em}}{2^{12}\pi^4}\,\frac{m_e^2\bfpe^2d\bfpe E_\nu' dE_\nu' }{(m_e+T_e)E_\nu M_A^2} \delta(E_\nu+m_e-\delta_B+M_A-E_\nu'-(m_e+T_e)-M_A)
|\langle p_1'p_2' p_3'|T|p_1p_2p_3\rangle |^2\,x\label{eq:Master}
} 
where $\delta_B$ is the binding energy and $T_e$ is the kinetic energy of the final electron $e'$, and $x$ is given in  Appendix \ref{AppA} for the process in consideration.
The following scalar products will be useful, 
\dis{
& q^2=-2E_\nu E_\nu'(1-\cos\theta),~q=k_\nu-k_\nu',\\
&k_\nu\cdot k_\nu'=-k_\nu\cdot q=k_\nu'\cdot q=-\frac{q^2}{2},\\
&p_e\cdot p_e'=m_e^2-\frac{q^2}{2}~\textrm{assuming atom at rest},\\
&k_\nu'\cdot p_e=m_e E_\nu'=k_\nu\cdot p_e'=E_\nu E_e'-E_\nu \bfpe\cos\theta_e, \label{eq:kinvar}
}
where we used massless neutrinos, $\theta_e$ is the polar angle of the outgoing  electron direction relative to the incoming neutrino direction,\footnote{The polar angle made by the outgoing  neutrino  direction relative to the incoming neutrino direction will be denoted as $\theta$.} and the non-relativistic approximation $E_e'\simeq m_e+\frac{\bfpe^2}{2m_e}=m_e+T_e$ is used. 

Firstly, let us show the inverse-$T_e$ rule, used in Sec. \ref{sec:XenonData}.
Note the kinematic variables for $\nu_e(k_\nu)+e(p_e)\to \nu_e'(k'_\nu)+e'(p'_e)$, for (almost) massless neutrinos and in the limit $M_A=\infty$,
\dis{
 &E_\nu'=|{\bf k}_\nu'|=\sqrt{E_e'^2+E_\nu^2-m_e^2},\\
&2k_\nu\cdot k_\nu'=2E_\nu E_\nu'(1-\cos\theta),
}
where $\theta$ is the angle of the outgoing netrino relative to the incoming neutrino. It is possible to express  $\theta$ in terms of $E_e'$ \cite{Vogel89},  $\cos\theta=(E_\nu+m_e)E_\nu^{-1}T_e^{1/2}(T_e+2m_e)^{-1/2}$ such that $|{\bf p}'_\nu|\simeq E_\nu'$ becomes
\dis{
E_\nu'\simeq \sqrt{E_\nu^2+2m_e T_e-2E_\nu \sqrt{2m_eT_e}(E_\nu+m_e)E_\nu^{-1}T_e^{1/2}(T_e+2m_e)^{-1/2}}.\label{eq:TeDep}
}
Now, the $|{\bf k}_\nu'|d|{\bf k}_\nu'|=E_\nu'dE_\nu'=-E_\nu' dT_e$ integration with the energy conservation delta function $\delta(f(T_e))$, with the approximation $T_e\ll m_e$ and $E_\nu'\simeq E_\nu$ in the final formula, we obtain
\dis{
\frac{d\sigma}{dT_e}\simeq \frac{1}{2^5\pi m_e^2 T_e  }. \label{eq:StiffPh}
}
The reason for the $1/T_e$ dependence is the specific $T_e$ powers in Eq. (\ref{eq:TeDep}), and if it is slightly violated then the $1/T_e$ dependence would not result. 
Thus, we anticipate the inverse-$T_e$ rule is not applicable if the momentum carried by the scattered Xenon atom is considered.

Let us now proceed the three-body scattering
given in Eq. (\ref{eq:3scattF}). The four momenta delta function 
\dis{
\delta^{(3)}({\bf k}_\nu-{\bf k}'_\nu-{\bf p}'_e-{\bf P}'_A) \delta(E_\nu-E_\nu'-\Delta-\frac{\bfpe^2}{2m_e}-\frac{\bfpa^2}{2M_A} )
}
where ${\bf p}_e' $ and  ${\bf P}_A' $ are the momenta of the outing $e'$ and outgoing atom $A'$ with $m_e$ and $M_A$, respectively, and  $E_\nu-E_\nu'= \Delta=T_e+\delta_B$ with the binding energy $ \delta_B$ of the electron in the atom.  The energy transfer to the atom is neglected in the energy-conservation $\delta$ function, since it is so small. The averaged spin sums of $|T|^2$ from Eq. (\ref{eq:Effatom}) is
\dis{
&\frac{G_F^2}{2M_{\rm eff}^{6}}{\rm Tr\,}\gamma^0(1+\gamma_5)\gamma^\rho\gamma^0\kslash' \gamma^\alpha(1+\gamma_5)\kslash\cdot
{\rm Tr\,} \gamma^0(g_V^*+g_A^*\gamma_5)\gamma_\rho\gamma_0(\pslash'+m_e)\gamma^0\gamma_\alpha(g_V+g_A\gamma_5)(\pslash+m_e)\cdot{\rm Tr\,}(\Pslash'+M_A)(\Pslash+M_A)\\
&~=\frac{G_F^2}{2M_{\rm eff}^{6}}{\rm Tr\,}2(1-\gamma_5)\gamma^0\gamma^\rho\gamma^0\kslash'\gamma^\alpha\kslash \cdot
{\rm Tr\,}\gamma^0\gamma_\rho\gamma^0(\pslash'+m_e)\gamma_\alpha\\
&\qquad\qquad\qquad\big[(|g_V|^2+|g_A|^2)\pslash- (g_Vg_A^*+g_V^*g_A)\pslash\gamma_5 +(|g_V|^2-|g_A|^2)m_e-(g_Vg_A^*-g_V^*g_A)m_e\gamma_5\big]\cdot(8M_A^2)\\
&~=\frac{G_F^2}{2M_{\rm eff}^{6}}2\cdot 4(k'^{\rho}k^\alpha -k\cdot k' g^{\rho\alpha}+ k'^{\alpha}k^\rho -i\varepsilon^{\rho\mu\alpha\nu}k'_\mu k_\nu) \times 4\big((|g_V|^2+|g_A|^2)(p'_\rho p_\alpha-p\cdot p'g_{\rho\alpha}+p'_\alpha p_\rho)\\
&\qquad~+(|g_V|^2-|g_A|^2)m_e^2g_{\rho\alpha}-i(g_Vg_A^*+g_V^*g_A)\varepsilon_{\rho\kappa\alpha\eta}p'^\kappa p^\eta
\big)\cdot(8M_A^2)\\
&~=\frac{G_F^2}{2M_{\rm eff}^{6}}2\cdot 4\Big(
4(|g_V|^2+|g_A|^2)(2k\cdot p k'\cdot p'+2k\cdot p' k'\cdot p)+4\big[(|g_V|^2-|g_A|^2)m_e^2(-2 k\cdot k')\\
&\qquad-2{\rm Re\,}(g_V g_A^*) (k\cdot p k'\cdot p'-k\cdot p' k'\cdot p)\big]
\Big)\cdot(8M_A^2)
\\
&~=\frac{2^8G_F^2M_A^2m_e^2 E_\nu^2}{M_{\rm eff}^{6}}\Big(
(|g_V|^2+|g_A|^2)\big[(1-\frac{\bfpe^2}{2m_eE_\nu}) +(\frac{E_e'}{m_e} -\frac{\bfpe}{m_e}\cos\theta_e)^2 \big]+ (|g_V|^2-|g_A|^2)m_e^2(-q^2/2E_\nu^2)\\
&\qquad-{\rm Re\,}(g_V g_A^*)\big[(1-\frac{\bfpe^2}{2m_eE_\nu}) +(\frac{E_e'}{m_e} -\frac{\bfpe}{m_e}\cos\theta_e)^2  \big]\Big)  \label{eq:Trace}
}
where we will use $M_{\rm eff}^6$ given in Eq. (\ref{eq:M6eff}) of Appendix \ref{AppA} and
  $\sum_s u_{e_{\rm atom}}(p)\bar{u}_{e_{\rm atom}}(p)\simeq\pslash+m_e$.  Here, however, the three momentum conservation is treated accurately. On the other hand, most people use the two body scattering and the uncertaintly  here is the probability in finding a prospective bound electron in the atom and the subsequent integration over the probability function of the electron cloud in the atom, which will be very complicated. This calculation is not found in the literature.

%%%%%%%%%%
\subsection{Magnetic moment contribution $|T|^2_{\rm MM}$}\label{subsec:MM}
If neutrinos have magnetic moment, we consider the second term in Eq. (\ref{eq:EMcoupling}). In this case, summing over the  spins, viz. Eq. (\ref{eq:Combine}), we have
\dis{
|T|^2_{\rm MM}&=\frac14\left( \frac{e^2 f_M}{2MM_{\rm eff}^3}\right)^2\frac{q_\mu q_\alpha}{q^4}\,{\rm Tr}\,\sigma^{\mu\nu}(\frac{\kslash_i}{2})\sigma^{\alpha\beta} (\frac{\kslash_f}{2})\cdot {\rm Tr}\,\gamma_\nu(\pslash_i+m_e)\gamma_\beta (\pslash_f
+m_e)\cdot{\rm Tr\,}(\Pslash'+M_A)(\Pslash+M_A)\\\\
&= \frac{1}{64}\left( \frac{e^2 f_M}{2MM_{\rm eff}^3}\right)^2\frac{1}{q^4}{\rm Tr} \,(\qslash\gamma^\nu-\gamma^\nu\qslash)\kslash_i(\qslash\gamma^x-\gamma^x\qslash) \kslash_f \cdot{\rm Tr}\gamma_\nu(\pslash_i+m_e)\gamma_\beta (\pslash_f
+m_e)\cdot (8M_A^2)\\
&= \frac{1}{64}\left( \frac{e^2 f_M}{2MM_{\rm eff}^3}\right)^2\frac{1}{q^4}\,Q^{\nu x}L_{\nu x}\cdot (8M_A^2)
}
where we neglected the neutrino mass. We obtain\footnote{The trace of  six gamma matrices, $
(1/4) {\rm Tr}\, \gamma^\rho \aslash  \bslash\gamma^\sigma \cslash\dslash= g^{\rho\sigma} (a\cdot b \, c\cdot d- a\cdot c \, b\cdot d+a\cdot d \, b\cdot c) +a\cdot c(b^\rho d^\sigma+b^\sigma d^\rho) 
+b\cdot d (a^\rho c^\sigma+a^\sigma c^\rho) \\
-a\cdot d(b^\rho c^\sigma+b^\sigma c^\rho) -b\cdot c(a^\rho d^\sigma+a^\sigma d^\rho) 
+a\cdot b(-c^\rho d^\sigma+c^\sigma d^\rho) +c\cdot d(a^\rho b^\sigma-a^\sigma b^\rho)
$,  is used.}
\dis{
Q^{\nu x} &\equiv {\rm Tr} (\qslash\gamma^\nu-\gamma^\nu\qslash)(\kslash_i+m_\nu)(\qslash\gamma^x-\gamma^x\qslash) (\kslash_f+m_\delta) \\
&=4g^{\nu x} (2 q\cdot k_i\, q\cdot k_f)+ 4k_i\cdot k_f  ( q^\nu q^x-q^2 \,g^{\nu x}) 
+4q^2(k_i^\nu k_f^x+k_i^x k_f^\nu ) \\
&\quad -4q^\nu( k_f^x\,q\cdot k_i+k_i^x\, q\cdot  k_f) -4 q^x k_f^\nu\,q\cdot k_i   +4q^x k_i^\nu\, q\cdot k_f , \\
L_{\nu x} &\equiv 4(p^i_\nu p^f_x+p^f_\nu p^i_x -p^i\cdot p^f g_{\nu x}+m_e^2 g_{\nu x}),\label{eq:Qnux}
}
from which  we obtain, neglecting O($m_e^2/E_\nu^2$),
 using Eq. (\ref{eq:kinvar}),
\dis{
\frac{1}{q^4}Q^{\nu x} L_{\nu x} =&\frac{1}{q^4}\Big( 32k_i\cdot k_f\,q\cdot p_i\, q\cdot p_f  
+16 q^2(-k_i\cdot k_f\, p_i\cdot p_f+ 2k_i\cdot p_i\, k_f\cdot p_f +2k_i\cdot p_f \, k_f\cdot p_i )\\
&-32q\cdot k_i (q\cdot p_i\, k_f\cdot p_f + q\cdot p_f\, k_f\cdot p_i)-32q\cdot k_f (q\cdot p_i\, k_i\cdot p_f + q\cdot p_f\, k_i\cdot p_i)\\
&+16 m_e^2(4 q\cdot k_i \,q\cdot k_f -q^2 k_i\cdot k_f ) \Big) \\
=&\frac{1}{q^4}\Big(-4q^6   +16 q^2( k_i\cdot p_i+ k_f\cdot p_i)(k_i\cdot p_f+k_f\cdot p_f)\Big)\\
& =4Q^2-\frac{16}{Q^2}(k_i\cdot p_i+k_f\cdot p_f)(k_i\cdot p_f+k_f\cdot p_i)=8E_\nu E_\nu'(1-\cos\theta)-\frac{8m_e^2}{ (1-\cos\theta)} \label{eq:MMterms}
}
where $q^2=-Q^2$, and $|T|^2_{\rm MM}$    is
\dis{
|T|^2_{\rm MM}=\frac{  \pi^2}{2}\left( \frac{ f_{\alpha\beta}}{m_e M_{\rm eff}^3}\right)^2\alpha^2_{\rm em} \Big(E_\nu E_\nu'(1-\cos\theta) -\frac{m_e^2}{ (1-\cos\theta)}\Big) (8M_A^2).\label{eq:MMspins}
 }
Using Eq. (\ref{eq:Master}), we obtain the cross section for the incident neutrino  $\nu_\alpha$ going to $\nu_\beta$ with a MM (including transition MM also), $f_{\alpha\beta}$ (in units of the electron Bohr magneton),\footnote{We use the incoherent cross sector for $\Delta p\gg 2.5\,\eV$.}  
\dis{
\frac{d\sigma}{d T_e d\cos\theta} &=\frac{3Z^3\alpha_{\rm em}^3m_e^3E_\nu' \bfpe}{2^{10}\pi^4 E_\nu (m_e+T_e)}
4\pi^2\left( \frac{ f_{\alpha\beta}}{m_eM_{\rm eff}^3 }\right)^2\alpha^2_{\rm em} \Big(E_\nu E_\nu'(1-\cos\theta) -\frac{m_e^2}{ (1-\cos\theta)}\Big)\\
&=\frac{3Z^3\alpha_{\rm em}^5m_e^3}{2^{8}\pi^2 M_{\rm eff}^6 }\frac{\sqrt{2T_e/m_e}}{ (1+T_e/m_e)^{5/2}} (1-\frac{\delta_B}{m_e}-\frac{T_e}{m_e})^2
\Big((1-\cos\theta) -\frac{m_e^2}{ E_\nu E_\nu'(1-\cos\theta)}\Big)\left( f_{\alpha\beta}\right)^2\\
&=0.883\times 10^{-5}\meV^{-3}\frac{\sqrt{2T_e/m_e}}{ (1+T_e/m_e)^{5/2}}(1-\frac{\delta_B}{m_e}-\frac{T_e}{m_e})^2
\Big((1-\cos\theta) -\frac{m_e^2}{ E_\nu E_\nu'(1-\cos\theta)}\Big)\left( f_{\alpha\beta}\right)^2\label{eq:Sc3MM}
}
leading to
\dis{
\frac{d\sigma}{d T_e}=0.883\times 10^{-5}\meV^{-3}\frac{\sqrt{2T_e/m_e}}{ (1+T_e/m_e)^{5/2}}(1-\frac{\delta_B}{m_e}-\frac{T_e}{m_e})^2
\Big(2 +\frac{2m_e^2}{ E_\nu E_\nu'}\ln\frac{1}{\delta_{\rm inf}}\Big)\left( f_{\alpha\beta}\right)^2
\label{eq:ScMM}
}
where  $\theta_{\rm min}$ is near 0 is used, \ie $\cos\theta|_{\rm max}=\frac12-\frac{\sqrt{2m_e T_e}}{<E_\nu'>}$,
which is  integrated over $T_e$ and the energy flux of solar $pp$ neutrinos \cite{Bahcall04} to give
\dis{
\textrm{For MM}:~\frac{d\sigma}{dT_e } 
&\simeq 1.071\times 10^{-5} \meV^{-3} \left( 2 -8.288 \right)f_{\alpha\beta}^2.\label{eq:sigMMnum}
}
The infrared divergence in the last term of Eq. (\ref{eq:Sc3MM}) is cured by the soft photon emission processes \cite{Bloch37}, and Eq. (\ref{eq:sigMMnum}) is used for $T_e\ge 2.5\keV$.  Equation (\ref{eq:sigMMnum}) gives a number $-6.73\times 10^{-5} \meV^{-3}f_{\alpha\beta}^2$.    Comparing with the almost flat central region ($T_e=15-20\,\keV$) of Fig.  \ref{fig:FittoEvents}\,(b), we obtain $|f_{\alpha\beta}|\le 0.86\times 10^{-7}$ which is much larger than that $2.8\times 10^{-11}$ obtained by the Borexino collaboration \cite{BdNuMagMom17}. The huge difference between our estimation and previous ones seems to be from the over estimation of cross sections in the previous analyses, using the inverse $T_e$ rule and  the coherent scattering.
   
%%%%%%%%%%
\subsection{Contributions from vector and axial-vector charges $|T|^2_Q$}

The electromagnetic $F_1^\nu$ form factor of a neutrino is identified from \cite{Giunti15}
\dis{
&\Lambda_\mu (q) =f_Q(q^2) \gamma_\mu -f_M(q^2) \,i\sigma_{\mu\nu}q^\nu +f_E(q^2)\sigma_{\mu\nu}q^\nu \gamma_5 +f_A(q^2)(q^2\gamma_\mu-q_\mu \qslash)\\
&F_1^\nu (q^2)=f_Q(q^2) =F_1^\nu (0) +q^2\frac{dF_1^\nu (q^2)}{dq^2}\Big|_{q^2=0}+\cdots.
} 
The charge radius is the $q^2$ dependent part,
\dis{
\tilde{r}^2\equiv  \langle r^2\rangle=6 \frac{dF_1^\nu (q^2)}{dq^2}\Big|_{q^2=0}.\label{eq:ChRad}
}
The elastic cross section includes the interference of the SM weak interaction and the vector and axial vector couplings of neutrinos with
\dis{G_V=\frac12+2\sin^2\theta_W+e^2\left(\frac{{\tilde r}^2/6}{\sqrt2 G_F}-\frac{\varepsilon}{\sqrt2 G_F Q^2}\right) , \textrm{ and} ~ g_A=\frac12,\label{eq:onGV}
} 
where $Q^2=-q^2>0$ for the $t$-channel scattering.
However, we consider the 3-body scattering to take into account the momentum of the recoiling atom. The parameter describing the atomic electromagnetic effects is described in Appendix \ref{AppA} in terms of $B$ and $C$. Inserting (\ref{eq:Trace}) into (\ref{eq:Master}), we obtain
\dis{
\frac{d\sigma}{dT_e} &=\frac{3Z^3\alpha_{\rm em}^3G_F^2 m_e^4  }{16\pi^4}\cdot \frac{ \sqrt{2m_eT_e}  E_\nu(E_\nu-\delta_B-T_e) }{M_{\rm eff}^{6}  (1+T_e/m_e)} \approx 1.22\times 10^{-23}\cdot\frac{ \sqrt{2m_eT_e}  E_\nu(E_\nu-\delta_B-T_e) }{  (1+T_e/m_e)}\cdot\frac{1}{M_{\rm eff}^{6}}\\
&=6.33\times 10^{-6}\meV^{-3} \cdot\frac{   \big[\frac{E_\nu}{m_e}(\frac{E_\nu}{m_e}-\frac{\delta_B+T_e}{m_e})\big] }{\big(1+\frac{T_e}{m_e}\big)^4}  \sqrt{\frac{2T_e}{m_e}}   \label{eq:SigDit}
} 
with $B\simeq 1.144\times 10^4$,  where the following $|T|^2_Q$ for the SM value \cite{tHooft71} is 
\dis{
|T|^2_Q=\frac{8G_F^2m_e^2}{ M_{\rm eff}^2} \,\left\{ (G_V+g_A)^2 E_\nu^2 +(G_V-g_A)^2 E_\nu E_\nu' +m_e E_e' (g_A^2-G_V^2)\right\},\label{eq:TQ2}
}
and  we used $1/M_{\rm eff}^6$ given in Eq. (\ref{eq:M6eff}).

%%%%%%%%%%
\subsection{Millicharge}\label{subsec:Mili}
  %%%%%%%%%%%%%%%%%%
\begin{figure}[!h]
 \includegraphics[height=0.25\textwidth]{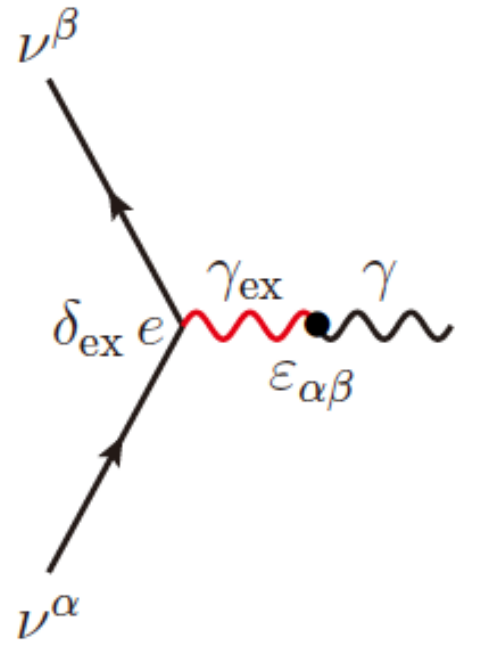}
 \caption{The neutrino coupling to photon through the kinetic mixing with ex-photon $\gamma_{\rm ex}$.}
\label{fig:ExPho}
\end{figure}
%%%%%%%%%%%%%%%% 

There can be another tiny EM property of neutrinos if there exist another (almost) massless U(1) gauge boson. If neutrinos couple to this ex-photon, $\gamma_{\rm ex}$, neutrinos can carry millicharges via the kinetic mixing \cite{Holdom86}, for which the Feynman diagram is shown in Fig. \ref{fig:ExPho}. Since the SM neutrino is in a SM doublet, L-handed electron $e_L$ must carry the same millicharge under the ex-photon gauge group.  Then, the magnitude of the $\nu_\mu$ coupling to the ex-photon is limited from the $\nu_\mu+e$ scattering data (through neutral current coupling by $Z_\mu$),
\dis{
(\delta_{\rm ex} e)_{\rm from\,\nu\,vertex}\frac{1}{q^2} (\delta_{\rm ex}e)_{\rm from\,e\,vertex}\le \frac{1.167\times 10^{-5}}{\sqrt2 \gev^2} \varepsilon(e_L)=0.413\times 10^{-5}/\gev^2
}
where $ \varepsilon(e_L)(=\varepsilon(\nu_L))$ is the neutral current coupling of $e_L$ to the $Z$ boson in the SM \cite{KimRMP81}, \ie $0.5$. 
Since there is no way to pinpoint $ \delta_{\rm ex}$, we set it as 1, transfering its uncertainty to the kinetic mixing value $\varepsilon=F_1^{\nu}(0)$.

%%%%%%%%%%
\section{Fitting}\label{sec:Fitting}
  
 The contributions from the MM and CR of neutrinos add incoherently. For the MM contribution, we obtained already Eq. (\ref{eq:ScMM}). For the CR contribution, we should consider the interference with the weak amplitude in the SM, which is given in Eq. (\ref{eq:SigDit}).   From Fig. \ref{fig:FittoEvents}\,(b), the cross section is less than $ 5\times 10^{-19}  \,\meV^{-3}$ for $T_e>17$\,keV. For $\sin^2\theta_W\simeq 0.231$ \cite{sin2LHC18,KimRMP81}, Eq. (\ref{eq:SigDit}) gives,  
  with $\langle E\rangle =0.2668$ and  $\langle E^2\rangle = 0.0788\meV^2$  of solar $pp$ neutrinos, at ${\tilde r}^2=\varepsilon =0$, 
\dis{
\textrm{For CR}: ~
\simeq  & 6.33\times 10^{-6} \meV^{-3}\cdot \Big\{\big(0.3018-1.02\, T_{e,\rm MeV}\big)\Big[1.462+ \Gamma\Big]^2\\
&+\big(0.3018-1.02\, T_{e,\rm MeV} \big)\Big[0.462+  \Gamma\Big]^2\Big\}\frac{2.768\, T_{e,\rm MeV}^{1/2} }{\big(1+1.957\, T_{e,\rm MeV} \big)^4} ,\label{eq:sigQnum}
}
where $T_{e,\rm MeV}=T_e/\meV$, which becomes at $T_e=17$\,keV  
\dis{
 & 0.5703\times 10^{-6} \meV^{-3}\cdot \Big\{\Big[1.462+ \Gamma\Big]^2 +\Big[0.462+  \Gamma\Big]^2\Big\} \to 1.341\times 10^{-6} \meV^{-3}{\rm ~for~}\Gamma=0.\label{eq:Gammais0}
} 
 The ratio of $ 5\times 10^{-19}  \,\meV^{-3}$ and (\ref{eq:Gammais0}) is $\approx 0.790\times 10^{-13}$. For solar neutrino energies of order 400\,keV, $Q^2$ given in  (\ref{eq:kinvar}) is of order $(0.3-0.4\,\meV)^2$. Taking $\langle Q^2\rangle\simeq 0.1\meV^2 $, the charge radius limit from Eq. (\ref{eq:ChRad}) is $2.18\times 10^{-6}\meV^{-1}$, or $4.30\times 10^{-17}$cm.
 Note that $\Gamma$ in Eq. (\ref{eq:sigQnum}) is
\dis{
\Gamma &=4\pi\alpha_{\rm em}\left(\frac{{\tilde r}^2/6}{\sqrt2 G_F}-\frac{\varepsilon}{\sqrt2 G_F Q^2}\right)= 0.9265\times 10^{9}\left( \tilde{r}^2\meV^2-6\varepsilon_{\rm MeV}\right)\\
&=2.387\times 10^{30}\left(  \tilde{r}^2{\rm cm}^{-2}-2.329\times 10^{-21}\varepsilon_{\rm MeV}\right),
}
with $\varepsilon_{\rm MeV}=\varepsilon/Q_{\rm MeV}^2$ (where $Q_{\rm MeV}$ is $Q$ in units of MeV).  

 %%%%%%%%%%%%%%%%%%
\begin{figure}[!t]
 \includegraphics[height=0.35\textwidth]{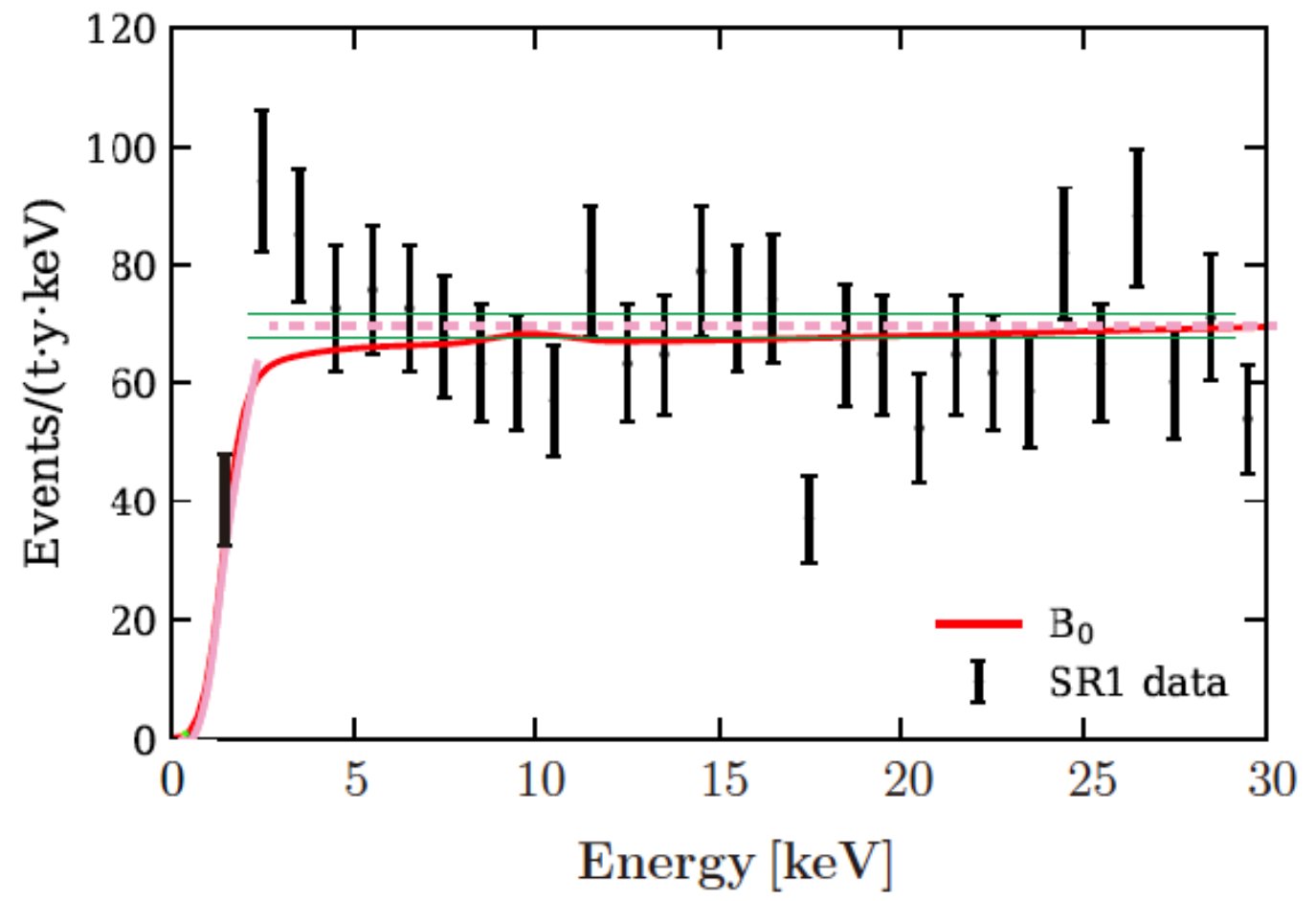}
 \caption{The green bars  are the  $1\sigma_{\rm stat}$-level backgrounds. The systematic error is  a factor 3 smaller than the statistical error \cite{Aprile20}.  }
\label{fig:Xenon1}
\end{figure}
%%%%%%%%%%%%%%%%

\begin{figure}[!h]
 \includegraphics[height=0.44\textwidth]{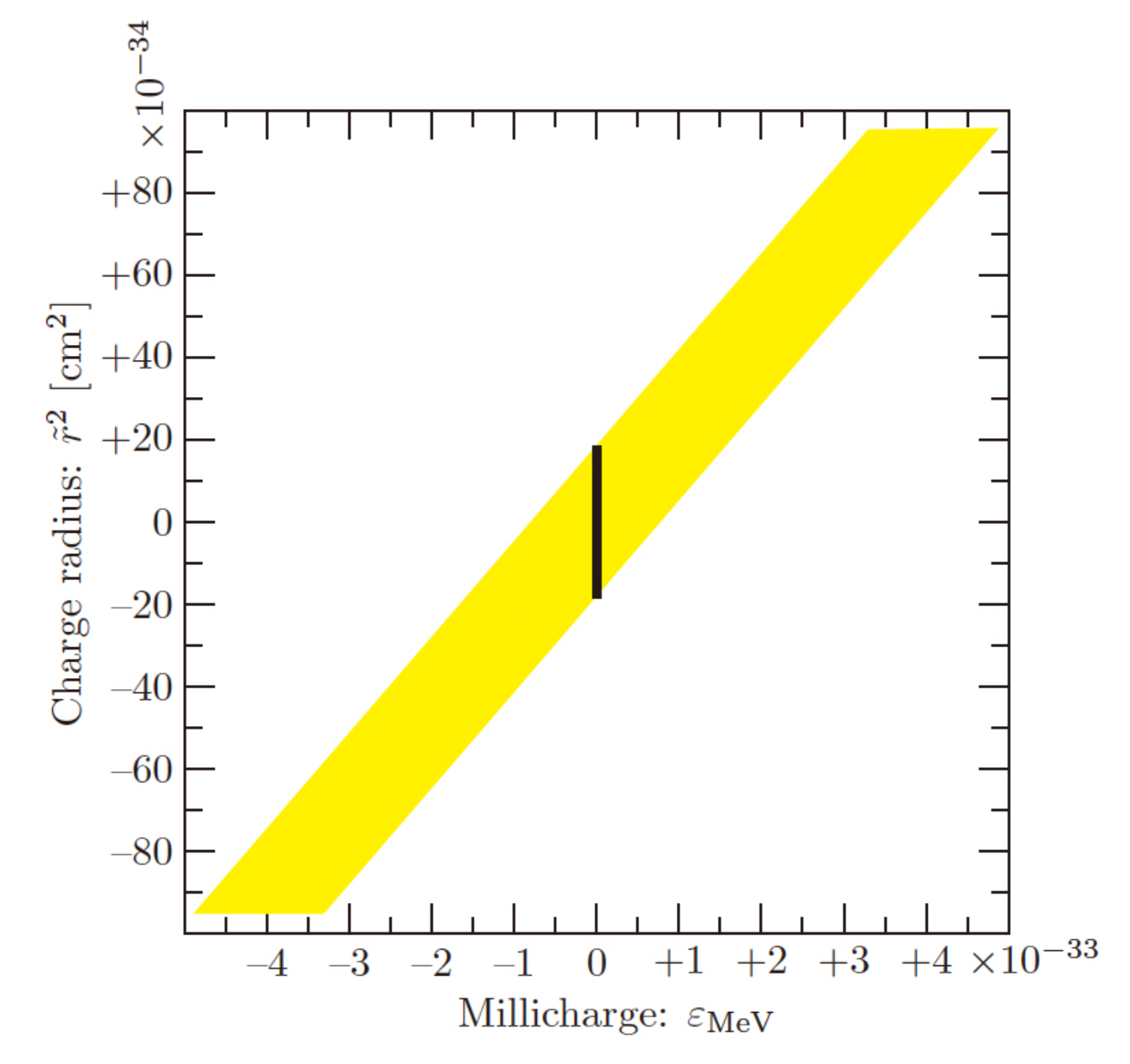}
\caption{The bounds on the charge radius (vertical bar) at $\varepsilon=0$ and the $1\sigma$ yellow band in the charge radius vs. millicharge plane, where $\varepsilon_{\rm MeV}=\varepsilon/Q_{\rm MeV}^2$ (where $Q_{\rm MeV}$ is $Q$ in units of MeV) depends on the effective momentum transfer in the process.}\label{fig:EMbounds}
\end{figure}

If we allow millicharges, the $1\sigma$ error (taken as the $\pm 2$ events out of 76 in Fig. \ref{fig:FittoEvents}\,(a))  is shown as the yellow band  in the $\tilde{r}^2-\varepsilon$ plane of Fig. \ref{fig:EMbounds}. There already exist bounds on the millicharges of light dark matter from the stellar evolution data, viz. Fig. 12 of 
\cite{CMBeps1} and Fig. 1 of  \cite{CMBeps2}, presenting $\varepsilon<10^{-12}$ in the dark matter mass range less than $10^4$\,eV, which however cannot be used for our SM neutrinos. The difference is that we studied in this paper  the (almost) massless SM neutrinos which cannot be dark matter in the Universe and we considered the EM properties of neutrinos only from solar neutrinos.
   %%%%%%%%%%%%%%%%%%
%%%%%%%%%%%%%%%%
    
%%%%%%₩96=87
\section{Conclusion}\label{sec:Conclusion}
We obtained the bounds on the electromagnetic properties of neutrinos implied by the XENON1T data:  a  bound on the magnetic moment  $|f_{\alpha\beta}|\le 0.86\times 10^{-7}$(times the electron Bohr magneton),  the charge radius $|\tilde{r}| < 4.30\times 10^{-17\,}{\rm cm}$, and the millicharge bound (Fig. \ref{fig:EMbounds}) if there exists a massless extra photon.   

%%%%%%%
\begin{appendix}
\section{Estimate of probability amplitude}\label{AppA}
If an electron is kicked out from an atomic orbit, an electron in the outer orbit will fill that vacancy immediately. The transition is occurring electromagnetically  via the $E1$ or $M1$ transition. The the $E1$ transition rate between the eigenstates $|k\rangle$ and $|n\rangle$ is presented a long time ago \cite{Schiff55,Gottfried03} 
\dis{
\frac{(2\pi)^2\alpha_{\rm em}}{3}\omega_{kn}^3|\langle k|{\bf r}|n\rangle|^2 ,
}
where ${\bf r}$ is the coordinate of the EM field. We can estimate $\langle  k|{\bf r}|n\rangle$ for the ground state ($k=n=1$)  as $ a_B/Z$,  with $a_B=0.529\times 10^{-8\,}$cm=268.1\,MeV$^{-1}$, and for the excited state $n$ as  $ na_B/Z$. We use $a_Z=14.48\keV^{-1}$ for $Z=54$. Then, the rest of the matrix element takes into account only the spherical harmonics and gives selection rules between angular momentum eigenfunctions. Now we consider kicking out the electron outside the atom and the selection rule always permits it. So, we will consider $\omega_{nn'}=\omega_{n\infty}=|E_{n}/\hbar|$.
Thus, the transition rate to kick out an electron to outside  is 
\cite{Gottfried03},
\dis{
\frac{4\alpha_{\rm em}k^3}{3} \frac{1}{2j_i+1}|\langle n_f j_f|| D  ||n_i j_i\rangle|^2\label{eq:ProbE1}
}
Considering the Hydrogenic radial wave function $C_n r^{n-1}e^{-r/na}$ with $C_n=(2/n)^{2n+1}/\sqrt{4\pi a_Z^3}$ with $a_Z=a_B/Z$, we can estimate the last factor, between the plane wave and the bound state $|n\rangle$,  $|2^{2n}(n+1)!a_Z^4/n^{n-1}\sqrt{\pi a_Z^3V}|^2$. Thus, the transition rate to kick out an electron to outside by electromagnetic interactions in the atom is  
\dis{
\frac{4\alpha_{\rm em}k^3}{3}\frac{a_Z^5}{\pi V}\left(\frac{2^{2n+1} [(n+1)!]}{n^{n-1}}\right)^2.
 \label{eq:PrE1}
}

Let us introduce a parameter $M_{\rm eff}$ reproducing Eq. (\ref{eq:PrE1}), using the following effective interaction  
\dis{
\frac{1}{M^2_{\rm eff}}\bar{e}\gamma^\mu e\bar{A}\gamma_\mu A,\label{eq:4ferInt}
}
and we use particle physicists'  $2\to n$  transition rate,
\dis{
Rate=(2\pi)^{4-3n}\frac{d^3{\bf p}_1'\cdots d^3{\bf p}_n' }{2E_12E_2 2E_1'2E_2'\cdots 2E_n'} \frac{\delta^4(p_1+p_2-p_1'\cdots-p_n')}{V}
|\langle p_1'\cdots p_n'|T|p_1p_2\rangle |^2.\label{eq:2nRate}
}
The two-body scattering rate from Eq. 
(\ref{eq:2nRate}), using (\ref{eq:4ferInt}) and averaging over a flux of incoming particles on a rest target,  is
\dis{
&(2\pi)^{-2}\frac{d^3{\bf p}_e'}{ 2^4m_eE_e'M_A^2}\frac{\delta(\sum_iE_i-\sum_fE_f)}{V} \frac{2^6}{M_{\rm eff}^4}m_eE_e'M_A^2\\
&=\frac{4\bfpe^2 d\bfpe }{}\frac{\delta( E_\nu'+m_e+\frac{\bfpe^2}{2m_e}+M_A-E_\nu -m_e+\delta_B-M_A)}{\pi V} \frac{1}{M_{\rm eff}^4}   
=4\pi^{-1}\frac{m_e\bfpe  }{M_{\rm eff}^4 V} 
}
where we used
\dis{ 
16(p_e^\mu {p'_e}^\nu -p'_e\cdot p_e g^{\mu\nu}+p_e^\nu {p'_e}^\mu+m_e^2 g^{\mu\nu})
(P_A^\mu {P'_A}_\nu -P'_A\cdot P_A g_{\mu\nu}+P_A^\nu {P'_A}_\mu+M_A^2 g_{\mu\nu})=\\
16(2p_e\cdot P_Ap_e'\cdot P_A' -2p'_e\cdot p_eP'_A\cdot P_A +2p_e\cdot P_A'p_e'\cdot P_A+2m_e^2P_A\cdot P_A' \\
-(P_A\cdot P_A')(-2p_e\cdot p_e'+4m_e^2) 
+M_A^2(-2p_e\cdot p_e'+4m_e^2)\\
\to\frac{32}{M_{\rm eff}^4} (p_e\cdot P_Ap_e'\cdot P_A'  +p_e\cdot P_A'p_e'\cdot P_A 
-M_A^2p_e\cdot p_e'+m_e^2M_A^2)\\
\simeq \frac{32}{M_{\rm eff}^2} \left(3 m_e^2M_A^2-m_e E_e'M_A^2\right)\simeq \frac{2^6}{M_{\rm eff}^4}m_eE_e'M_A^2 .
}
This can be compared to (\ref{eq:PrE1}), and  we obtain
\dis{
\frac{1}{M_{\rm eff}^2}=\sqrt{\frac{\alpha_{\rm em}}{3}}\frac{a_Z^{5/2}}{\sqrt{m_e (m_e+T_e)} } k_n^{3/2}\left(\frac{2^{2n+1} [(n+1)!]}{n^{n-1}}\right) ,\label{eq:Fix}
}
where we use $k_n$ for the threshold value, $k_n\simeq |E_n(Xe)|/\hbar$.
In practice, let us use Eq. (\ref{eq:PrE1}) summing over the electrons from each orbit of the Xenon atom with $k_n^3$,
\dis{
& A\equiv\sum_{n=1}^5\,\frac{1}{n^3}f(n)\left(\frac{2^{2n+1} [(n+1)!]}{n^{n-1}}\right)^2=7.028\times10^5,\\
& B\equiv\sum_{n=1}^5\,\frac{1}{n^6}f(n)\left(\frac{2^{2n+1} [(n+1)!]}{n^{n-1}}\right)^2\simeq 1.144\times 10^4,
\label{eq:Sumsn}
}
where $f(n)=2,8,18,18,8$ for $n=1,2,\cdots,5$, respectively. This takes into account the number of electrons in the Xenon atom $Z$, corresponding to the incoherent process. $B$ appears when we takes into account $V$ in the denominator, which is the case in the 3-body scattering.
 Thus, we estimate, for the Xenon atom of $Z=54$ and $M_A\simeq 7.028\times 10^5\meV$, 
\dis{
\frac{1}{M_{\rm eff}^2}=\sqrt{\frac{\alpha_{\rm em}}{3}}\frac{\sqrt{A}}{Z^{5/2}m_e^2\sqrt{1+T_e/m_e}} =0.150 (1+T_e/m_e)^{-1/2}\,\meV^{-2}  .\label{eq:FixMeff}
} 
If we use (\ref{eq:FixMeff}), then
\dis{
\frac{1}{M_{\rm eff}^6}=5.69\times 10^{4}\frac{\keV^{-15/2}}{[m_e(m_e+T_e)]^{3/2}} \sum_nk_n^{9/2}\left(\frac{2^{2n+1}[(n+1)!]}{n^{n-1}} \right)^3\equiv 2.27\times 10^{5}\frac{\meV^{-3}}{[m_e(m_e+T_e)]^{3/2}}\,C\label{eq:M6eff}
}
where $C$ becomes\footnote{If we use $Z_{\rm eff}(n)$, including electron interactions, we obtain $C=5.136\times 10^{12}$, and ${1}/{M_{\rm eff}^6}= 1.17\times 10^{18}[m_e(m_e+T_e)]^{-3/2}\,\meV^{-3}$. 
}
\dis{
C\simeq 2.287\times 10^{12}.
}

The $3\to n$  transition rate is
\dis{
Rate=(2\pi)^{4-3n}\frac{d^3{\bf p}_1'\cdots d^3{\bf p}_n' }{2E_12E_2 2E_32E_1'2E_2'\cdots 2E_n'} \frac{\delta^4(p_1+p_2+p_3-p_1'\cdots-p_n')}{V^2}
|\langle p_1'\cdots p_n'|T|p_1p_2p_3\rangle |^2\label{eq:3nRate}
}
from which we calculate the three body scattering cross section. A flux of incoming particles 1 (neutrino in our case) scatter on two particles 2 (electron in our case) and 3 (atom in our case). Thus, the cross section consists of two parts, one with a flux factor $1/|{\bf v}_1-{\bf v}_2|$ and the other $1/|{\bf v}_1-{\bf v}_3|$.
Taking ${\bf v}_3=0$, the neutrino-electron scattering is
\dis{
d\sigma=\frac{(2\pi)^{4-9}}{|{\bf v}_1-{\bf v}_2|}\frac{d^3{\bf p}_1'd^3{\bf p}_2'  d^3{\bf p}_3' }{2E_12E_2 2E_32E_1'2E_2'2E_3'} \frac{\delta^4(\sum p_i-\sum p_f')}{V}
|\langle p_1'p_2' p_3'|T|p_1p_2p_3\rangle |^2\label{eq:3scattF}
}
where we can take $V$ as the atomic volume to which the flux of neutrinos sweep. For $1/V$, the sum $B$ in Eq. (\ref{eq:Sumsn}) will be used below. The $T$-matrix squared in Eq. (\ref{eq:3scattF}) is summed over the spins
\dis{
\frac{1}{(2s_e+1)(2s_\nu+1)}\sum_{s_\nu,s_e}\sum_{s'_\nu,s'_e}|{\rm Amp}|^2=  (|T|^2_{\rm Q}).
}
If we consider the MM interaction also, then we will have a factor  $(|T|^2_{\rm MM} +|T|^2_{\rm Q})$ instead of $(|T|^2_{\rm Q})$. For the volume $V$, we use the Xenon volume for each $n$ separately. The $n$-th shell volume we used $V=(4\pi /3)(na_Z)^3$. 
\end{appendix}
 
%%%%%%%%
\acknowledgments{\noindent We thank Kyungwhan Ahn, Elena Aprile, Ki-Young Choi, Paul Frampton, Wonho Jhe,  Junu Jeong, Sin-Kyu Kang,  Sejin Kim, and Myungbo Shim   for useful discussions.  J.E.K. thanks the APCTP, for the hospitality extended to him during his visit, where this work has been started. J.E.K. is supported in part  by the National Research Foundation (NRF) grant  NRF-2018R1A2A3074631, and S.Y. is supported in part by the Institute for Basic
Science (IBS-R017-D1-2020-a00/IBS-R017-Y1-2020-
345 a00).}

 %%%%%%%%%%%%%%%%%%%%%%%%% 
%\newpage
  %%%%%%%%%%%%%%%%%%  

  \end{document}